\documentclass[reprint,aps,prl,
]{revtex4-2}
\usepackage{amsmath}
\usepackage{amsfonts}
\usepackage{amssymb}
\usepackage{mathrsfs}
\usepackage{bbm}
\usepackage{graphicx}
\usepackage{dcolumn}
\usepackage{bm}
\usepackage{hyperref}
\usepackage{pstricks}
\usepackage{pst-3dplot}
\usepackage{subfigure}
\usepackage{xy}

\newpsobject{showgrid}{psgrid}{subgriddiv=1,griddots=10,gridlabels=6pt,gridcolor=green}

\definecolor{darkblue}{rgb}{0,0,.7}
\hypersetup{colorlinks=true, breaklinks=true, linkcolor=darkblue, menucolor=darkblue, urlcolor=darkblue, citecolor=darkblue}

\begin{document}

\title{A gravitational metrological triangle}

\author{Claus L\"ammerzahl$^1$}
 \email{claus.laemmerzahl@zarm.uni-bremen.de}
 \altaffiliation[Also at ]{Gauss-Olbers Space Technology Transfer Center, c/o ZARM, University of Bremen, 28359 Bremen, Germany}
\author{Sebastian Ulbricht$^{2,3}$}%
 \email{sebastian.ulbricht@ptb.de}
\affiliation{$^1$Center of Applied Space Technology and Microgravity ZARM, University of Bremen, Am Fallturm, 28359 Bremen, Germany \\
$^2$Fundamentale Physik für Metrologie FPM, Physikalisch-Technische Bundesanstalt PTB, Bundesallee 100, 38116 Braunschweig, Germany\\
$^3$Institut für Mathematische Physik, Technische Universität Braunschweig, Mendelssohnstraße 3, 38106 Braunschweig, Germany}

\date{February 5, 2024}

\begin{abstract}
Motivated by the similarity of the mathematical structure of Einstein's General Relativity in its weak field limit and of Maxwell's theory of electrodynamics it is shown that there are gravitational analogues of the Josephson effect and the quantum Hall effect. These effects can be combined  to derive a gravitational analogue of the quantum/electric metrological triangle. 
The gravitational metrological triangle may have applications in metrology and could be used to investigate the relation of the Planck constant to fundamental particle masses. This allows for quantum tests of the Weak Equivalence Principle. Moreover, the similarity of the gravitational and the quantum/electrical metrological triangle can be used to test the universality of quantum mechanics.

\end{abstract}

\maketitle

\section{Metrology, the new SI, and metrological triangles}
A major aspect of metrology is the definition and dissemination of physical units such as the SI base units second, meter, kilogram, coulomb, kelvin, mol, and candela \cite{SIBrochure}. 
While once these units were based on specific realizations and artefacts, they are now defined through the fixed numerical values of seven fundamental constants. Besides the second as a given number of oscillations of light, emitted from a certain atomic transition in Caesium, those constants are the velocity of light $c$, the Planck constant $h$, the electric charge $e$, the Boltzmann constant $k_B$, the Avogadro constant $N_A$, and the luminous efficacy $K_{\mathrm{cd}}$ \cite{Mills2006}, each playing central roles in various branches of physics. It is clear that, by this choice, the Meter Convention established a deep relation between metrology and the structure of physics and Special and General Relativity, in particular \cite{Hehl:2018gen}.

\begin{figure}[b]
\includegraphics[width=\columnwidth]{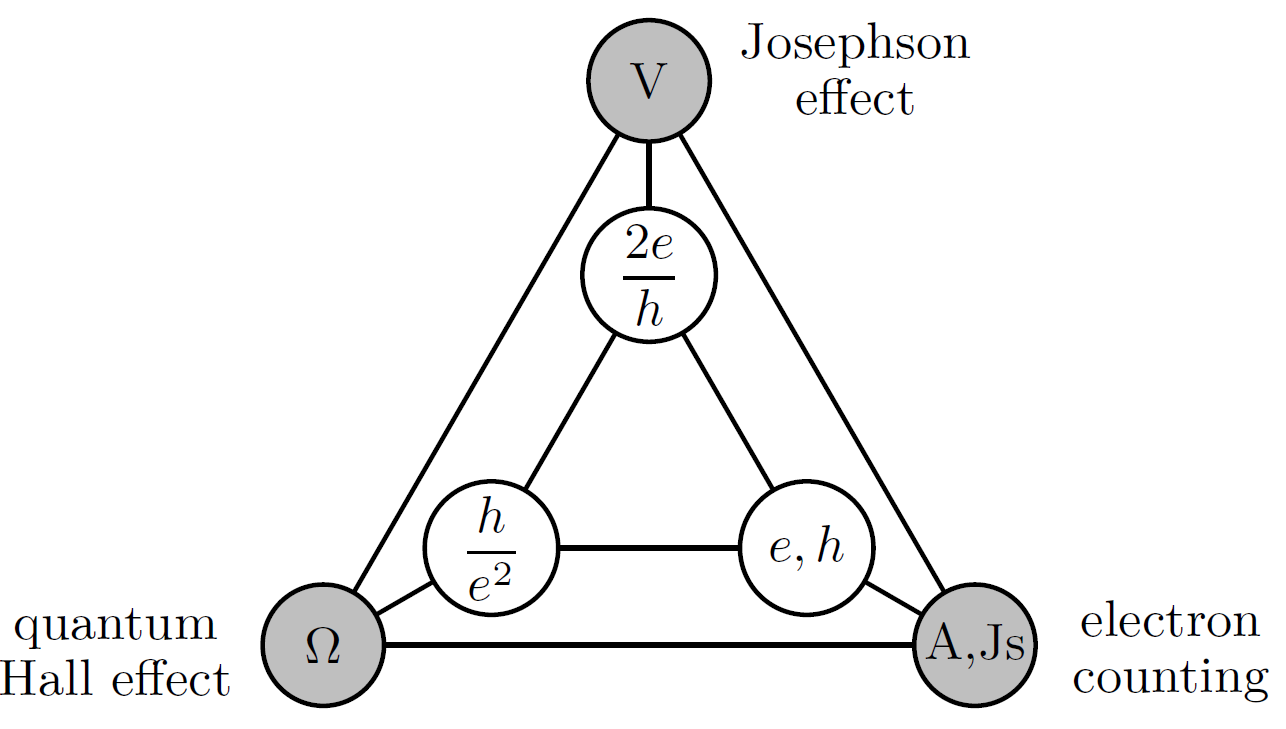}
\caption{The metrological triangle for $e$ as well as for $h$. \label{eMT}}
\end{figure}

Before the redefinition of the SI was put into practice in 2019, it was needed to measure the defining constants to highest precision \cite{Liebisch2019,BIPM2007,RevModPhys.93.025010}. In the case of the electron charge $e$ and the Planck constant $h$ this only became possible thanks to the quantum revolution in metrology, heralded by the discovery of the Josephson effect and the quantum Hall effect \cite{Josephson1962,Klitzing1980}. Each of these effects gave experimental access to a fundamental measurand, the Josephson constant $K_{\text{J}} = 2e/h$ and the von Klitzing constant $R_{\text{K}} = h/e^2$, respectively. Together with the electron charge $e$, these constants form the \emph{quantum/electrical metrological triangle} \cite{PecolaetalRMP2013,SchereretaAdP2019}, see Fig.~\ref{eMT}, where the value of $e$ can be obtained from electron counting, e.g. using a single electron pump. This framework allowed for different schemes to measure $e$ and $h$, and thus offered substantial options for consistency checks improving our understanding of the underlying physical processes. Also today, after the SI redefinition, the quantum/electrical metrological triangle remains a powerful tool for metrology and fundamental physics. 

Nowadays, the constituting effects of the triangle are used to \emph{realize} electric standards, such as the units ampere, ohm, and volt which complement the realizations of the SI base units. For instance, an array of Josephson junctions is used to realize a voltage of $\pm 10\,\mathrm{V}$ with only  $\mathrm{nV}$ uncertainty \cite{Behr2012,Tang2012}. That voltage could be used further to obtain a precise current from a quantum Hall device. The same device also can be used to compare resistors with the exact value of the von Klitzing constant $R_{\text{K}} =25.8128...\,\mathrm{k}\Omega$. Thus, the individual effects, but also their interplay can be exploited for scientific and technical application. Based on these unit realizations the effects can be exploited for the investigation of electromagnetic fields, for instance by measuring the induced voltage in a conductor or the magnetic flux in a SQUID. This also allows for a precise and independent determination and detailed study of additional fundamental constants, such as the vacuum permittivity $\epsilon_0$ and vacuum permeability $\mu_0$, which should be related by $\epsilon_0\mu_0=1/c^2$. From these, further quantities, e.g., the wave impedance of the vacuum  $Z_0 = \sqrt{\mu_0/\epsilon_0}\approx 377\, \Omega$ and the fine structure constant $\alpha = e^2(4 \pi \epsilon_0 \hbar c)^{-1} = Z_0(2 R_\text{K})^{-1}\approx 1/137$ can be derived and measured in optical and atomic physics experiments.

In this letter we will propose to follow the same line of thought to introduce a \emph{quantum/gravitational} metrological triangle. For that we use the fact that Einstein's General Relativity in its weak field limit resembles the mathematical structure of electrodynamics. We deduce the gravitational analogues of the Josephson and the quantum Hall effect, which together with the counting of elementary particle masses, form a triangle that provides a great potential to investigate the interaction of gravity and quantum particles in various experiments.

\section{The structure of weak field Einstein gravity}

In what follows, we will motivate 
that gravity, in its weak field limit, is described by similar equations as electrodynamics while both are fundamentally different in nature.
Einstein's General Relativity is based on the Einstein field equation 
\begin{equation}
R_{\mu\nu} - \frac12 g_{\mu\nu} R = \frac{8\pi G}{c^4} T_{\mu\nu} \, .
\end{equation}
Here, $R_{\mu\nu}$ and $R$ are the Ricci tensor and Ricci scalar, respectively, and $g_{\mu\nu}$ is the space-time metric. $T_{\mu\nu}$ is the energy  momentum tensor and $G$ the Newton gravitational constant. A standard weak field approximation uses a decomposition of the metric into a Minkowskian background $\eta_{\mu\nu}$ and a small variation $h_{\mu\nu} \ll 1$
\begin{equation}
g_{\mu\nu} = \eta_{\mu\nu} + h_{\mu\nu} \, . 
\end{equation}
If we define the mass density $\rho := \frac{1}{c^2} T^{tt}$, the mass flux density $j^i := \frac{1}{c^2} T^{ti}$, and the gravitational scalar and vector potentials 
\begin{equation}
\phi := \bar h^{tt} c^2 \, , \qquad  a^i := \bar h^{ti}    \, , 
\end{equation} 
where $\bar h_{\mu\nu} = h_{\mu\nu} - \frac12 \eta_{\mu\nu} \eta^{\rho\sigma} h_{\rho\sigma}$ \cite{Schafer:2008jea}, then we obtain the linearized Einstein equations in harmonic gauge 
\begin{align}
\pmb{\nabla} \cdot {\pmb{B}}_{\text{g}} & = 0 & \quad \pmb{\nabla} \times {\pmb{E}}_{\text{g}} + {\dot{\pmb{B}}}_{\text{g}} & = 0 \label{GRM1} \\
\pmb{\nabla} \cdot {\pmb{E}}_{\text{g}} & = 4 \pi G \rho & \quad \pmb{\nabla} \times {\pmb{B}}_{\text{g}} - \frac{1}{c^2} {\dot{\pmb{E}}}_{\text{g}} & = \frac{4 \pi G}{c^2} \pmb{j} \label{GRM2}
\end{align}
where we defined the weak field gravitational field strengths, that is, the gravitoelectric and the gravitomagnetic field strengths
\begin{equation}
	{\boldsymbol{E}}_{\text{g}} = - \boldsymbol{\nabla} \phi - \dot{\boldsymbol{a}} \, , \quad {\boldsymbol{B}}_{\text{g}} = \boldsymbol{\nabla} \times \boldsymbol{a} \, .
\end{equation}
Boldface symbols are 3-vectors and the dot denotes the time derivative. With that, \eqref{GRM1} and \eqref{GRM2} are the weak field Einstein equations, see e.g. \cite{Schafer:2008jea,CiufoliniWheeler1995}. 

We may push even further the analogy to the Maxwell equations by eliminating the velocity of light: Introducing another gravitational constant $\mu_{\text{g}} := c^2/G$ allows us to define the gravitoelectric and gravitomagnetic excitations ${\boldsymbol{D}}_{\text{g}} := \frac{1}{G} {\boldsymbol{E}}_{\text{g}}$ and ${\boldsymbol{H}}_{\text{g}} := \mu_{\text{g}} {\boldsymbol{B}}_{\text{g}}$ and to rewrite the inhomogeneous equations \eqref{GRM2} as
\begin{align}
\boldsymbol{\nabla} \cdot {\boldsymbol{D}}_{\text{g}} & = 4 \pi \rho & \quad \boldsymbol{\nabla} \times {\boldsymbol{H}}_{\text{g}} - {\dot{\boldsymbol{D}}}_{\text{g}} & = 4 \pi \boldsymbol{j} \, . \label{weakgrav2}
\end{align}
Having formulated the weak field Einstein equations \eqref{GRM1} and \eqref{weakgrav2} without the speed of light  we obtained on this level a pre-metric version of these equations \cite{Hehl:2016glb}. The constant $1/G$ now plays the role of a gravitoelectric vacuum permittivity, and $1/\mu_{\text{g}}$ the role of a gravitomagnetic vacuum permeability. $G$ gives the strength of the gravitoelectic attraction between two masses and $\mu_{\text{g}}$ gives the strength of the gravitomagnetic interaction between spinning masses. Accordingly, one can use test particles to measure the gravitoelectric attraction generated by given masses as well as the gravitomagnetic effects generated by rotating masses such as the Lense-Thirring or the Schiff effect, from which one can determine the two constants $G$ and $\mu_{\text{g}}$. Both constants together give a velocity which usually is interpreted as the velocity of light $c$, that is, $G \mu_{\text{g}} = c^2$. However, in principle we should interpret this $c$ as the speed of gravity, $G \mu_{\text{g}} = c_{\text{g}}^2$. In this way, that is, by measuring $\epsilon_0$ and $\mu_0$ from the motion of charged particles and $G$ and $\mu_{\text{g}}$ from the motion of massive particles, we have formulated a genuine test of whether the speed of gravity coincides with the speed of light. -- As in the electromagnetic case we also may define a gravitational wave impedance of vacuum $Z_0^{(\text{g})} := \sqrt{G/\mu_{\text{g}}}$ which is $\sim 2.34 \cdot 10^{-19}\;\text{m}^2/\text{kg}\,\text{s}$. The unit $\text{m}^2/\text{kg}\,\text{s}$ is the mechanical analogy to the unit $\Omega$ of the electrical resistance in the relation $I = U/R$, if $I$ describes a mass current in kg/s and $U$ a gravitational potential difference in $\text{m}^2/\text{s}^2$.

Having discussed the strong analogy between electrodynamics and weak field gravity regarding the field equations as well as the equations of motion of charged and massive particles, we now ask for implications of this analogy for massive quantum particles. 

\section{A gravitational Josephson effect}

The analogy between electrostatic and Newtonian gravitational forces gives a motivation to formulate a gravitational analogue of the electric Josephson effect.
For a gravitational Josephson effect we first need a gravitational Josephson junction. This can be realized using quantum condensates on a horizontal table \cite{WallisDalibardCohenTannoudji1992,Abele:2012dn} with a step of height $h_0$ and a barrier, giving a non-symmetric double well potential (DWP). In the following we consider neutral quantum particles. 

\begin{figure}[t]
\includegraphics[width=\columnwidth]{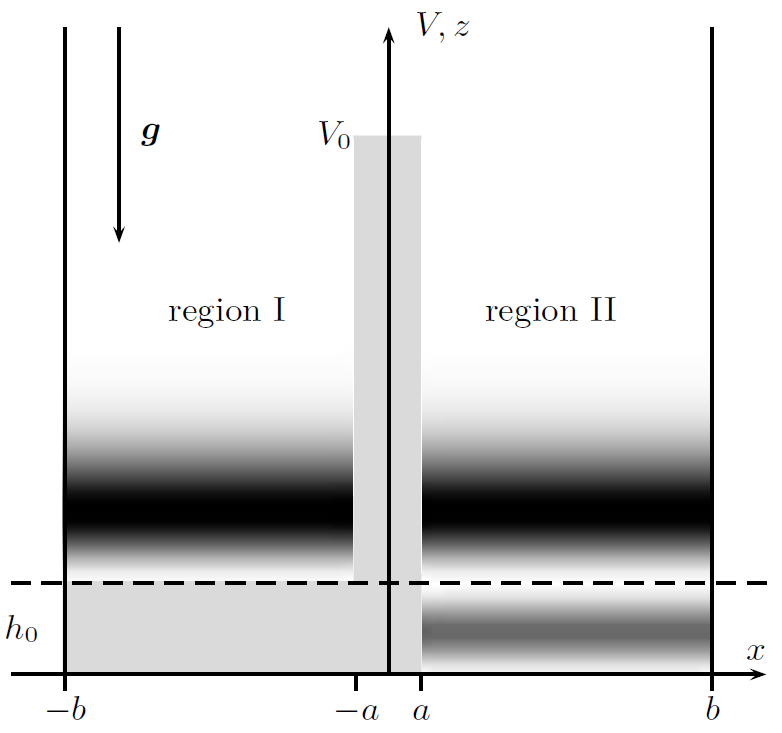}
\caption{The choice of the height $h_0$. The potential is shown in gray, the Airy function is given in terms of their position probability. The horizontal dashed line indicates the node of the Airy function which coincides with the height $h_0$ of the left region. $2a$ is the width of the tunnel barrier $V_0$ and $2b$ is the size of the box. (The $z$-axis shows the density of the wave function depending on $z$ but also shows the height of the potential $V_0$.) \label{Airyglueing}}
\end{figure}

The non-symmetric DWP we are considering is a box of width $2b$, a step of height $h_0$ at $x=0$ and a barrier of finite height and of width $2a$, see Fig~\ref{Airyglueing}. The whole DWP is in a constant gravitational field. The step provides a constant potential difference $g h_0$ between two regions where $g$ is the constant gravitational acceleration. The total Hamiltonian of the system is given by $H=H_0+V(x)$ with 
\begin{equation}
H_0 =  -\frac{\hbar^2}{2m}\Delta + m g z + \begin{cases} V_0 & \text{for $|x| < a$} \\ 0 & \text{elsewhere} \end{cases} \, , \label{H0}
\end{equation}
describing a symmetric DWP with a tunnel barrier of height $V_0$, and a small potential step $h_0$ making the DWP non-symmetric
\begin{equation}
V(x) = \begin{cases} m g h_0 & \text{for $x < -a$} \\ 0 & \text{elsewhere} \end{cases} \, . \label{PotStep} 
\end{equation}
We will solve this problem whereby the non-symmetric part will be treated as perturbation. 

We have a stationary situation so that $H \psi = E \psi$ with $E > 0$. Since in the considered problem the $y$ dependence is trivial we will omit this degree of freedom in the following. Then the above problem is a problem in $x$ and $z$. We assume $\psi(x, z) = \psi_x(x) \psi_z(z)$ and take $\psi_x(x) = e^{\pm i k x}$ where the $k$ is allowed to differ in the regions I,II and in the barrier region m. Inserting this separation ansatz in the Schr\"odinger equation we obtain the following three equations
\begin{eqnarray}
\left(E - m g h_0 - \frac{\hbar^2 k_{\text{I}}^2}{2m}\right) \psi_{z\text{I}} & = & - \frac{\hbar^2}{2m} \psi_{z\text{I}}^{\prime\prime} + m g z\, \psi_{z\text{I}}  \nonumber \\
\left(E - V_0 - \frac{\hbar^2 k_{\text{m}}^2}{2m}\right) \psi_{z\text{m}} & = & - \frac{\hbar^2}{2m} \psi_{z\text{m}}^{\prime\prime} + m g z\, \psi_{z\text{m}} \label{Schr1}\\
\left(E - \frac{\hbar^2 k_{\text{II}}^2}{2m}\right) \psi_{z\text{II}} & = & - \frac{\hbar^2}{2m} \psi_{z\text{II}}^{\prime\prime} + m g z\, \psi_{z\text{II}} \nonumber \, , 
\end{eqnarray}
where the prime denotes a derivative with respect to $z$. See also Fig.~\ref{Airyglueing}. A global solution to these equations is given by 
\begin{equation}
\psi_{z}(z) \sim \text{Ai}\left(\frac{z}{\ell} - \epsilon_i\right) \,   \label{eqn:wave_functions}
\end{equation} 
that reduces to $\psi_{z\text{I}}$, $\psi_{z\text{m}}$ and $\psi_{z\text{II}}$ in the corresponding domains.
 Here $\ell = \left(2 m^2 g/\hbar^2\right)^{- 1/3}$ is a characteristic length scale and $\epsilon_i$ is a zero of the Airy function, see e.g. \cite{WallisDalibardCohenTannoudji1992}. The energy and the momenta take the values $E=\frac{\hbar^2k^2}{2m}+mg\ell \epsilon_i$, $k_{\text{II}}\equiv k$, $k_{\text{I}}=\sqrt{k^2-2m^2gh_0/\hbar^2}$ and $k_{\text{m}}=\sqrt{2mV_0/\hbar^2-k^2}\equiv \kappa$.

Now we have to incorporate the boundary conditions regarding the coordinate $z$: $\psi_z(h_0) = 0$, and $\psi_z(0) = 0$. One way to achieve this is to choose $h_0=(\epsilon_i-\epsilon_j)\ell$ for any two zeros $i>j$ of the Airy function. The boundary and jump conditions regarding the coordinate $x$ leads to the quantization of the energy. The corresponding energies and wave functions can be determined along the usual procedure, see the Supplement Materials. The non-symmetric DWP potential then is implemented as perturbation. This way we obtain the Josephson equations
\begin{eqnarray}
i \hbar \dot w_{\text{L}} & = & E_{\text{L}} w_{\text{L}} + K w_{\text{R}} \\
i \hbar \dot w_{\text{R}} & = & E_{\text{R}} w_{\text{R}} + K w_{\text{L}} \, ,
\end{eqnarray}
characterizing the tunneling dynamics in a non-symmetric DWP. Here, the $w_{\text{L,R}}$ are the probabilities to find the particle in the region I or II. $E_{\text{L,R}}$ is the energy of the corresponding states, and $K = \frac12 \Delta E$ with $\Delta E = m g h_0$.  

We obtained equations which have exactly the same structure as for the description of the Josephson effect in superconductivity. Thus, we obtain the same solutions, that is, for a constant $h_0 \neq 0$ we get the AC Josephson effect with a current $I = I_0 \sin(\nu t)$ with $\nu = \frac{m}{h} v_0$ where $v_0  = g h_0$. For a mass of 1 amu and $h = 1\;\text{cm}$ we get $\nu = 10^4\;\text{Hz}$ and for $h \sim 10\;\mu\text{m}$ we have $\nu \sim 10\;\text{Hz}$. Based on that we can define a \textit{gravitational Josephson constant}
\begin{eqnarray}
K_{\text{J}}^{\text{(g)}} := \frac{m}{h} \qquad \text{in} \qquad \frac{\text{s}}{{\text{m}}^2} = \frac{\text{Hz}}{\frac{\text{m}^2}{\text{s}^2}} \, .
\end{eqnarray}
For neutrons, for instance, we get $K_{\text{J}}^{\text{(g)}} = 1.2667 \cdot 10^6 \frac{\text{s}}{{\text{m}}^2}$. In order to measure this effect one needs a collective quantum system as, e.g., a Bose-Einstein condensate. For a system of incoherent particles the phases are different and average out so that no net current will be observable. In fact, the Josephson constant for BECs in a non-symmetric DWP has been measured in \cite{Albiezetal:2005,Levietal2007}. 

\begin{figure}[t]
\includegraphics[width=\columnwidth]{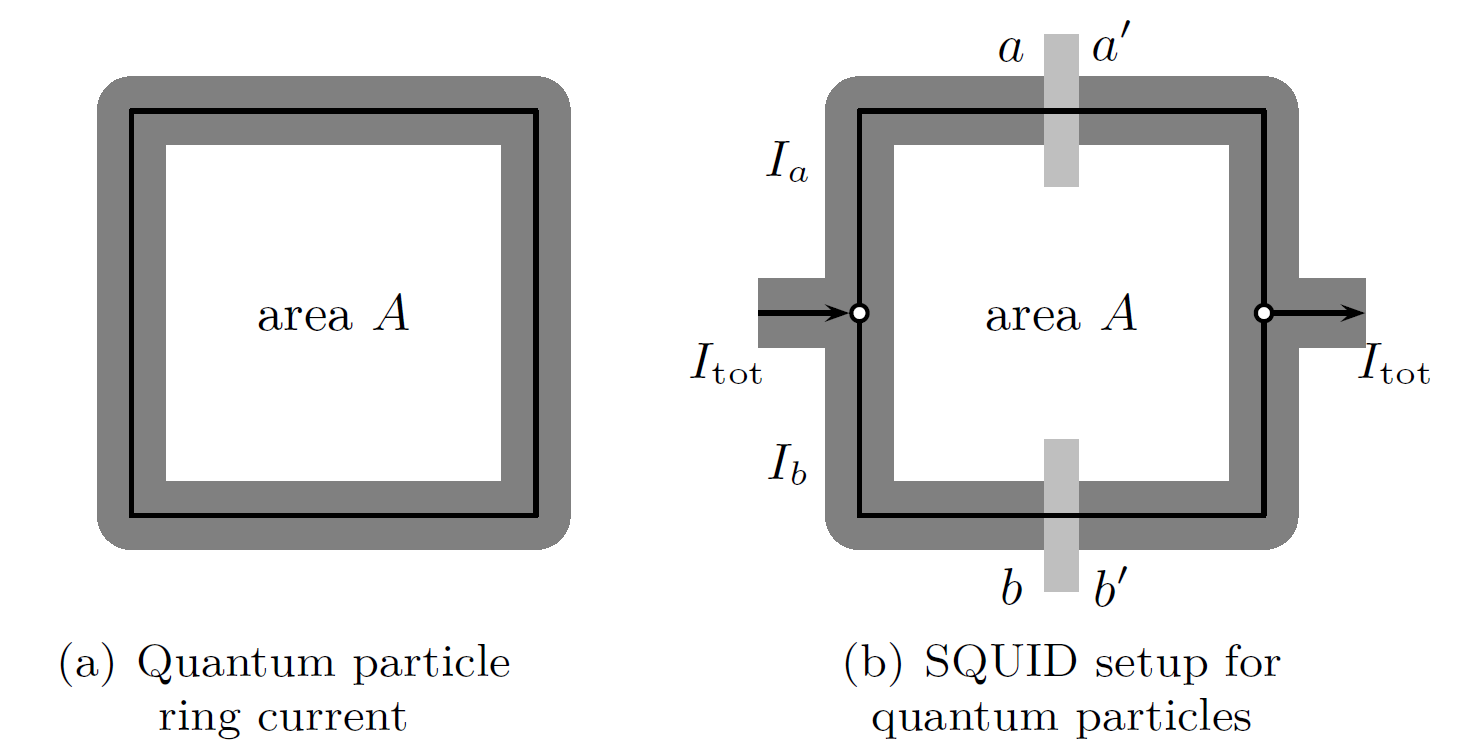}
\caption{Quantum particle SQUID for measuring rotations and gravitomagnetism. \label{grSQUID}}
\end{figure}

This Josephson junction may be used to build up a gravitational SQUID, see Fig.~\ref{grSQUID}(a). For any ring quantum particle current in a rotating frame using $\boldsymbol{k} = \boldsymbol{v} + \boldsymbol{\omega} \times \boldsymbol{x}$ we have
\begin{equation}
2\pi n =\oint \boldsymbol{k} \cdot d\boldsymbol{x} = \frac{m}{\hbar} \oint \pmb{v} \cdot d\pmb{x} + \frac{m}{\hbar} \Phi^{(g)} \approx \frac{m}{\hbar} \Phi^{(g)}
\end{equation}
for small velocities. Based on that we can define a flux quantum $\Phi_0^{(g)} := \frac{h}{ m}$ so that
\begin{equation}
\Phi^{(g)} = n \Phi_0^{(g)} \, . 
\end{equation}
This is deeply related to the Sagnac effect. The rotation $\boldsymbol{\omega}$ may consist of a kinematical rotation and of the gravitomagnetic rotation.  

In analogy to the electromagnetic case the current in Fig.~\ref{grSQUID}(b) is given by 
\begin{equation}
I_{\text{tot}} = 2 I_c \sin\delta \cos\left(\pi \frac{\Phi^{(g)}}{\Phi_0^{(g)}}\right)
\end{equation}
with the critical current $I_c$ and with 
\begin{equation}
\delta = \gamma_a + \pi \frac{\Phi^{(g)}}{\Phi_0^{(g)}} 
\end{equation}
where $\gamma_a$ is the phase shift at the Josephson junction $a$, Fig.~\ref{grSQUID}(b). The typical resolution of the flux is $\sim 10^{-6} \Phi_0^{(g)}$. For an area of $\Sigma \sim 10^{-2} \; {\text{m}}^2$ this implies a sensitivity of $\omega \sim 10^{-10} \; \text{Hz}$ for measuring a rotation. This is a gyroscope based on the Sagnac effect for matter waves realized through a gravitational SQUID for BECs, see also \cite{Levietal2007}.  

\section{A gravitational quantum Hall effect}

\begin{figure}[t]
\includegraphics[width=\columnwidth]{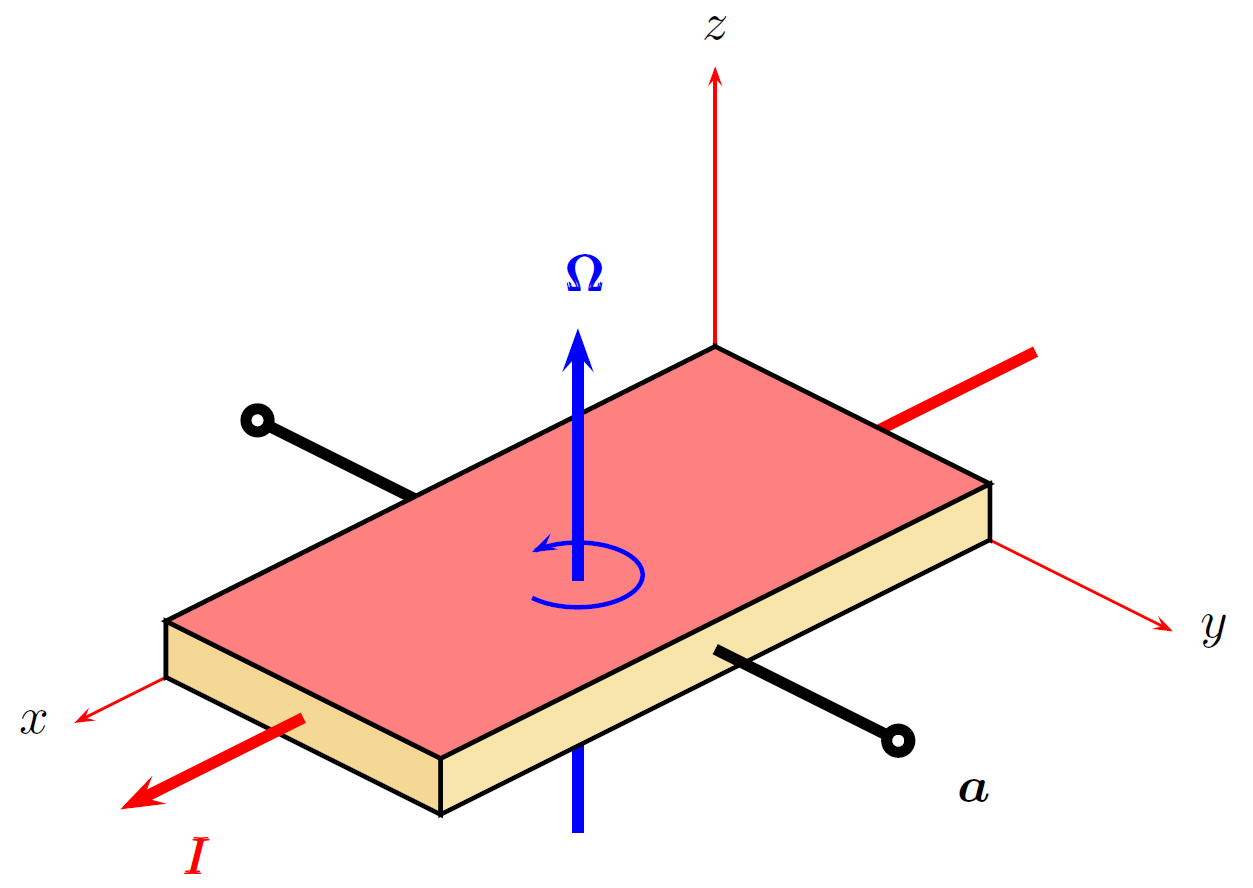}
\caption{Setup for a gravitational quantum Hall effect. Neutrons or atoms are moving on a thin rotating plate with a current $I$. The rotation induces an acceleration field orthogonal to the motion of the particles. A compensation of the transversal acceleration can be achieved by a small tilt of the plane around the $x$-axis what induces a small acceleration $\boldsymbol{a}$. \label{gravQH}}
\end{figure}

With mass currents, e.g. neutrons, atoms, or BECs, on a rotating surface a gravitational analogue of the quantum Hall effect can be realized. According to Fig.~\ref{gravQH}, in thin rotating plate the incoming current (red) will be deviated to the right. This can be compensated by a small gravitational acceleration in $y$-direction induced by a slight tilting of the $x-y$-plane in $y$-direction.

The Hamilton operator of a point mass in a rotating frame \cite{KlinkWickramasekara2013} is given by
\begin{equation}
H = \frac{1}{2m} {\boldsymbol{p}}^2 + \boldsymbol{\omega} \cdot \boldsymbol{L} \, ,
\end{equation}
where $\boldsymbol{L}$ is the angular momentum of the particle. If the rotation is around the $z$-axis and the particles move only in $x$-$y$-plane, see Fig.~\ref{gravQH}, then $p_z = 0$ and the Hamilton operator reads
\begin{eqnarray}
H & = & \frac{1}{2m} {\pmb{p}}^2 - \omega_z y p_x \\
& \approx & \frac{1}{2m} \left(p_x - m \omega_z y\right)^2 + \frac{1}{2m} p_y^2 \, ,
\end{eqnarray}
where we neglected terms quadratic in the rotation. This Hamiltonian has exactly the same structure as the Hamilton operator for the quantum Hall effect in the Landau gauge. One can calculate the gravitational analogue of the Landau levels and the mass current for the case that the Landau levels are filled up to some level $s$. By tilting the plane one may also induce a small gravitational acceleration needed to compensate the rotation induced acceleration, see Fig.~\ref{gravQH}. This then gives a gravitational quantum Hall effect described by the transversal mass current resistivity
\begin{equation}
\rho_{xy} = \frac{h}{m^2} \frac{1}{s} \, .
\end{equation}
This leads to the \textit{gravitational von Klitzing constant}
\begin{equation}
R_{\text{K}}^{(\text{g})}: = \frac{h}{m^2} \qquad \text{in} \qquad \frac{{\text{m}}^2}{\text{kg}\;\text{s}}
\end{equation}
for the used masses. If we take for instance neutrons or hydrogen atoms, then $R_{\text{K}}^{(\text{g})} = 2.481 \cdot 10^{18} \; \frac{{\text{m}}^2}{\text{kg}\;\text{s}}$. 

The classical gravitational Hall effect adds the Coriolis and gravitational acceleration $\boldsymbol{a} = \boldsymbol{g} + \boldsymbol{v} \times \boldsymbol{\Omega}$ with $\boldsymbol{v}=\boldsymbol{p}/m$, and requires $\boldsymbol{a} = 0$, $\boldsymbol{\Omega} = \Omega_z {\boldsymbol{e}}_z$ and $\boldsymbol{v} \bot \boldsymbol{\Omega}$. In equilibrium we have $v_y = 0$ so that
$\pmb{g} = v_x \Omega_z {\pmb{e}}_y$. Accordingly, the induced acceleration can be compensated by the plane being slightly tilted in $y$-direction (for $\Omega = 0.1 \;\text{Hz}$ and $v_x = 10\;\text{cm/s}$ we need $g = 1\;\text{cm/s}^2 = 10^{-3} g_{\text{Earth}}$ what requires a tilt angle of $\sim 1'$). 

This can also be described within the Drude model. The equation of motion including a friction/damping term for reaching an equilibrium ($\tau$ is a damping time) then is $\boldsymbol{a} = \boldsymbol{g} + \boldsymbol{v} \times \boldsymbol{\Omega} - \frac{\boldsymbol{v}}{\tau}$. In equilibrium $\boldsymbol{a} = 0$ and with $\boldsymbol{j} = n m \boldsymbol{v}$ we get Ohm's law $\boldsymbol{g} = \rho \boldsymbol{j}$ with the resistivity
\begin{equation}
\rho = \begin{pmatrix} \frac{1}{n m \tau} & - \frac{\Omega_z}{n m} \\ \frac{\Omega_z}{n m} & \frac{1}{n m \tau} \end{pmatrix} \, ,
\end{equation}
from which we can read off the longitudinal and transversal (Hall) resistivity
\begin{equation}
\rho_{xx} = \frac{1}{n m \tau} \qquad \rho_{xy} = \frac{\Omega_z}{n m} \, . 
\end{equation}
\section{A gravitational metrological triangle}
\begin{figure}[t]
\includegraphics[width=\columnwidth]{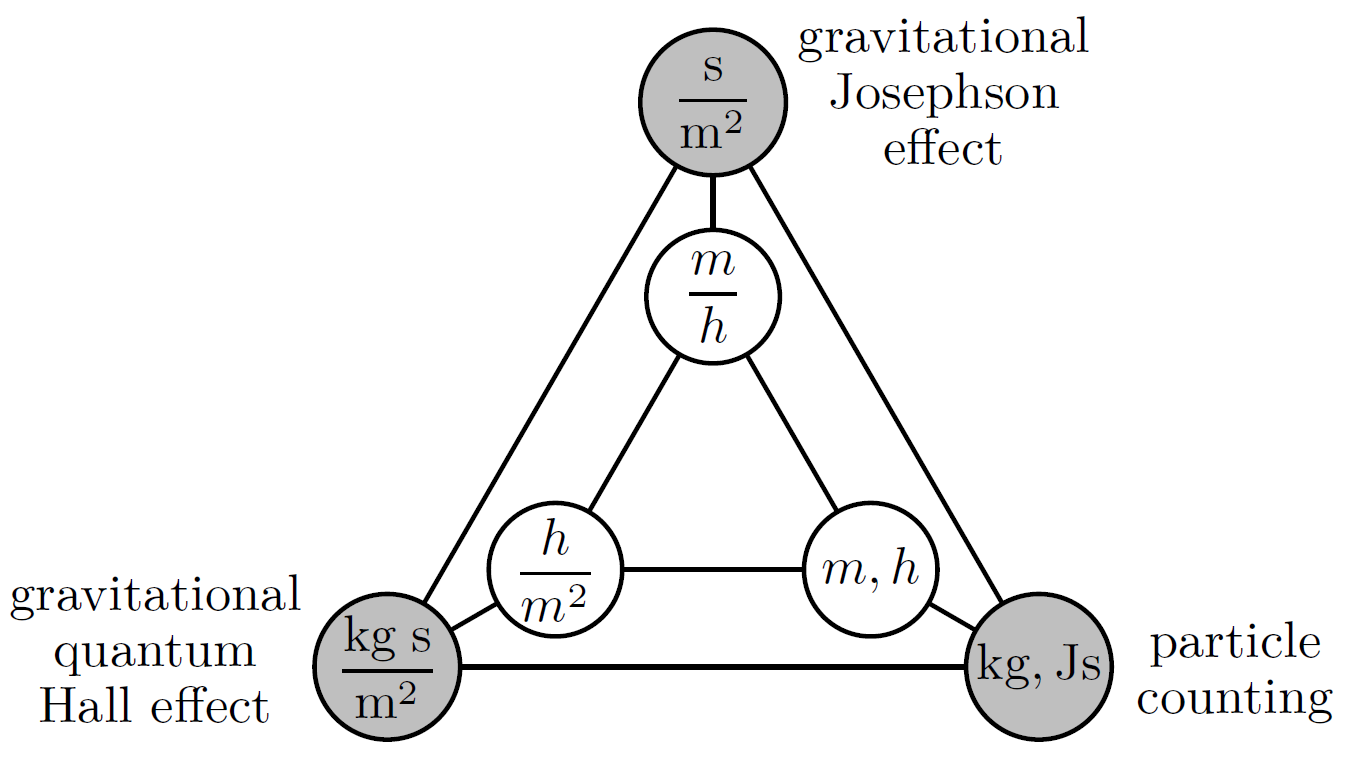}
\caption{The metrological triangle for $m$ as well as for $h$.\label{gMT}}
\end{figure}

Having discussed the gravitational Josephson effect and the gravitational quantum Hall effect, we can use these results to construct the gravitational analogue of the quantum/electrical metrological triangle.
In total analogy to the electric constants $e$, $K_{\text{J}}$, and $R_{\text{K}}$, which obey the relations \cite{PecolaetalRMP2013,SchereretaAdP2019}
\begin{eqnarray}
e = \frac{2}{K_{\text{J}} R_{\text{K}}} \quad &,& \quad h = \frac{4}{K_{\text{J}}^2 R_{\text{K}}}\label{eleceandh}\,,
\end{eqnarray}
we have found their gravitational counterparts, which satisfy 
\begin{eqnarray}
m = \frac{1}{K_{\text{J}}^{(\text{g})} R_{\text{K}}^{(\text{g})}}\quad &,&\quad h = \frac{1}{K_{\text{J}}^{(\text{g})2}R_{\text{K}}^{(\text{g})}} \, , \label{hmfromKJandRK}
\end{eqnarray}
see Fig.~\ref{gMT}.
Here, we find that the mass $m$ appears on the same level as $2e$ in (\ref{eleceandh}), where it represents the charge of a Cooper pair.
In (\ref{hmfromKJandRK}), however, all effects needed to realize the corresponding experiments are purely gravitational and without electromagnetic interaction.  At one hand, this could open a new window to quantum based metrology. At the other hand, this relations give raise to a number of possible consistency checks, to test our understanding of quantum systems in gravity. Ignoring for a moment, that we fixed $h$ in the current SI, we could assume experiments, where a fundamental mass $m$ is precisely known. In this version of the gravitational metrological triangle, we would have access to three measurands  
\begin{equation}
h \, , \qquad m/K_{\text{J}}^{(\text{g})}\, , \qquad m^2/R_{\text{K}}^{(\text{g})} \, , \label{Triangleh}
\end{equation}
which have to coincide. Within the modern SI, where the Planck constant $h$ is exactly known we, in turn, have three methods to obtain the mass of a chosen reference particle, 
\begin{equation}
m \, , \qquad  K_{\text{J}}^{(\text{g})} h \, , \qquad \sqrt{R_{\text{K}}^{(\text{g})} h} \label{Trianglem}\,,
\end{equation}
which allows similar insights as in the previous case. In that scenario, however, we have an additional option, since the fixed constant $h$ is the same as for quantum/electrical metrological triangle. Assuming the Plank constant to have the same value for the electromagnetic \emph{and} the gravitational interaction of quantum particles, we can relate its identities from Eqs. (\ref{eleceandh}) and (\ref{hmfromKJandRK}) to obtain  
\begin{equation}
 K_{\text{J}}^2 R_{\text{K}}=4 K_{\text{J}}^{(\text{g})2} R_{\text{K}}^{(\text{g})}  \, .
\end{equation}
Thus, we find both triangles to be not independent, what allows to check whether the Planck constant $h$ is the same in the gravitational and in the electrical context.
This might be interpreted as a test for the universality of quantum mechanics \cite{Fischbach:1991eg}.

\section{Summary and discussion}

A gravitational analogue of the quantum/electrical metrological triangle has been introduced. The new triangle is based on quantum effects of neutrons or atoms in a gravitational field -- in complete analogy to the electromagnetic interaction of charged quantum particles used in the quantum/electrical triangle. This is possible since the mathematical structure of weak field gravity resembles the governing equations of electromagnetism. We showed that two fundamental quantum effects can also be realized with the gravitational interaction: the gravitational Josephson effect and the gravitational quantum Hall effect. Together with particle counting in analogy to the count of electrons in a single electron pump those can be used to trace back for instance mass measurements to the values of the Planck constant $h$ and a chosen atomic or fundamental particle mass $m$. Fundamental masses can be determined to tremendous precision, for instance in particle trap experiments \cite{RevModPhys.93.025010,Blaum2006}. With that a pure quantum realization of the kilogram -- besides the Avogadro experiment \cite{Fujii2016} and the Kibble balance \cite{Robinson2016} -- may be possible, see also \cite{CCM2021}.

In addition to their metrological aspects, the gravitational Josephson effect and the gravitational quantum Hall effect can also provide fundamental test of the behavior of quantum systems in gravity: They are suitable to construct tests of the universality of quantum mechanics, i.e, whether quantum mechanics is valid and the same for interactions of quantum particles with electromagnetic and gravitational fields. According to \cite{Albiezetal:2005,Levietal2007}, the gravitational Josephson constant might be measurable with an accuracy of $10^{-3}$. The gravi\-tational quantum Hall effect has not yet been experimentally realized. 

As in the electromagnetic case one may use devices based on the discussed effects for a better exploration of the dynamics of matter and, thus, of the gravitational fields. However, the new approach will not help in improving the precision of measurements due to the weakness of gravity on the laboratory scale. Owing to the Weak Equivalence Principle, the mass of a test particle plays no role in the dynamics of \textit{test} particles, and in the astrophysical context a mass $M$ appears in the combination $GM$ only. 

Nevertheless, from the motion of test objects it is possible to measure the gravitational constants $G$ and $\mu_{\text{g}}$. The Newton gravitational constant $G$ is given by the force between two massive bodies \cite{RevModPhys.93.025010}, and the gravitomagnetic constant $\mu_{\text{g}}$ can be determined from the Lense-Thirring effect \cite{Ciufolini:2016ntr}. $G$ can be measured with 0.1\% precision, and for  $\mu_{\text{g}}$ one approaches 1\%. In analogy to the electromagnetic case one may -- in principle -- measure $G$ with clocks in a gravitational field given by a laboratory mass or measure the gravitomagnetic field with gravitational SQUIDs. 

A further issue is a hypothetical time dependence of the constants. While the SI-defining constants are constant by definition, the measured constants like $\epsilon_0$, $\mu_0$, or $G$ (or parameters relevant for the standard model of elementary particles) may depend on time. While a time dependent $G$ will not interfere with the constancy of the defining constants a time-dependent $\epsilon_0$, for example, will influence the energy levels of atoms and, thus, the definition of the second. However, since the second is based on one transition only, nothing will change and everything still is consistent. One may detect a hypothetical time-dependence of $\epsilon_0$ -- what is equivalent to a time dependence of the fine structure constant -- through the comparison of different energy transitions within one atom but this does not influence the particular transition chosen for the definition of the second. In any case, any hypothetical time dependence of $G$ or the fine structure constant has been experimentally excluded with highest precision \cite{Hofmann:2018myc,Lange:2020cul}.

From $G$ and $\mu_{\text{g}}$  further quantities can be derived, e.g., the gravitational wave impedance $Z_0^{(\text{g})} := \sqrt{G/\mu_{\text{g}}}\sim 2.34 \cdot 10^{-19}\text{m}^2/\text{kg}\,\text{s}$, which governs the propagation of gravitational waves in vacuum. 

Having started our investigation at the level of Einstein equations, we also have to take the Weak Equivalence Principle as granted. That means that one has to be sure that the gravitational interaction acts universally, that is, independent from the composition and weight of pointlike bodies. 
If we distinguish between inertial and gravitational mass, then inspection of the derivation of the gravitational Josephson and quantum Hall effects shows that the gravitational Josephson constant depends on the gravitational mass $K_{\text{J}}^{(g)} = \frac{ m_{\text{g}}}{h}$ and the gravitational von Klitzing constant on the product of inertial and gravitational mass $R_{\text{K}}^{(g)} = \frac{h}{m_{\text{i}} m_{\text{g}}}$. Then in the relations \eqref{hmfromKJandRK} the left hand sides will be replaced by $h \rightarrow h \frac{m_{\text{i}}}{m_{\text{g}}}$ and $m \rightarrow m_{\text{i}}$, respectively. This agrees with the general observation that in all equations of motion of test matter the gravitational mass appears in the combination $m_{\text{g}}/m_{\text{i}}$. Thus, in principle the MTs also provide a pure quantum test of the Weak Equivalence Principle when realized with different types of, e.g., atoms. The recently published final result of the MICROSCOPE mission provided a classical test of the Weak Equivalence Principle confirming it at the order $10^{-15}$ \cite{MICROSCOPE:2022doy}, see also \cite{Singh:2022wyp}. 

\begin{acknowledgments}
We thank H. Abele, G. Czycholl, F. W. Hehl, T. Mehlst\"aubler, A. Pelster, G. Sch\"afer, A. Surzhykov, and J. Ullrich for fruitful discussions. We acknowledge the support by the Deutsche Forschungsgemeinschaft (DFG, German Research Foundation) under Germany’s Excellence Strategy-EXC-2123 “QuantumFrontiers” -- Grant No. 390837967 and the CRC 1464 “Relativistic and Quantum-based Geodesy” (TerraQ). 
\end{acknowledgments}


%

\section{Supplementary material}

Here we derive the essential equations for the gravitational Josephson effect. For that we first determine the energies and eigenstates for the DWP and then the dynamics of symmetric and antisymmetric states in a non-symmetric DWP. 

\subsection{The symmetric double well potential}

We take the symmetric DWP as given by \eqref{H0}. We employ the boundary conditions in $x$-direction, that is, $\psi_x(\pm b)=0$ and the smoothness of the wave function at $x=\pm a$. In what follows, we first consider the symmetric potential ($h_0=0$) and treat the asymmetry as a perturbation afterwards. Then the wave functions are
\begin{equation}
	\psi_x(x) = \begin{cases} A \sin\left(k\left(x + b\right)\right) & \!\!\!\!\!\!\!\!\!\text{for } -b < x < - a    \\
		A \frac{e^{\kappa a} \sin\left(k (b - a)\right)}{s e^{\kappa a} + e^{- \kappa a}}  \\ 
		\qquad \times \left(e^{\kappa (x - a)} + s e^{- \kappa (x + a)}\right) & \text{for } |x| < a \\
		A s \sin\left(k \left(- x + b\right)\right)  & \text{for } a < x < b \, , \end{cases}
\end{equation}
where $s = \pm 1$ stands for the symmetric/antisymmetric solution, respectively. $A$ is a normalization factor and $\kappa$ and $k$ have to obey the condition
\begin{equation}
	\kappa \tanh^s\left(\kappa a\right) = - k \cot \left(k (b - a)\right) \, ,
\end{equation}
what leads to a discrete set of solutions for $k$ and $\kappa$. For simplicity, we consider the case of weak coupling between I and II, that is, $\kappa a \gg 1$, and of low energies of the particles, $\frac{k}{\kappa} \ll 1$. We then obtain the momentum and energy eigenvalues
\begin{align}
	k_n &= \frac{1}{b - a} \left(n \pi - \frac{k_n^0}{\kappa_n^0} \left(1 \pm e^{- 2 \kappa_n^0 a}\right)\right)  \\
	E_{i,n,\pm} &= E_n^0 \left(1 - \frac{2}{(b - a) \kappa_n^0} \left(1 \pm 2 e^{- 2 \kappa_n^0 a}\right)\right) + m g \ell \epsilon_i\, .
\end{align}
with $\kappa_n^0 = \frac{1}{\hbar} \sqrt{2m(V_0 - E_n^0)}$, $k_n^0 = \frac{n \pi}{b - a}$, and $E_n^0 = (\hbar k_n^0)^2/(2m)$ where $n$ should be not too large. We also define the average energy and the energy difference
\begin{eqnarray}
	E_{i,n}&=& \frac12 (E_{i,n,+}+E_{i,n,-}) \\
	\Delta E_n &=& E_{i,n,+} - E_{i,n,-} = \frac{8 E_n^0}{(b - a) \kappa_n^0} e^{- 2 \kappa_n^0 a} 
\end{eqnarray}
of symmetric and anti-symmetric states of given $n$ and $i$.

\subsection{The gravitational Josephson equations for the non-symmetric DWP}

Having solved the problem for the symmetric DWP, we now investigate the influence of the small potential step $gh_0$. Application of the standard first order perturbation theory to the small potential step \eqref{PotStep} yields the corrected states
\begin{equation}
	\tilde\psi_{i,n,s}(x,z) = \psi_{i,n,s}(x,z) + \frac{s}{2} \frac{m g h_0}{\Delta E_n}\, \psi_{i,n,-s}(x,z) \label{eqn:corrected_states}
\end{equation}
obeying to first order approximation
\begin{equation}
	[H_0 + V(x)] \tilde\psi_{i,n,s}(x,z) = \left(E_{i,n,s} + \frac12 m g h_0\right) \tilde\psi_{i,n,s}(x,z) \, .
\end{equation}
Thus, the small potential step $gh_0$ leads to a mixing between symmetric $(s=+1)$ and anti-symmetric $ (s=-1)$ wave functions, while mixing between states of different $n$ is suppressed by a factor $\Delta E_n/E_{i,n,s}$. In what follows we drop the indices $n$ and $i$ assuming them to be fixed and show them only when needed.

We now want to define left and right bin states by a superposition of eigen states of the non-symmetric DWP (\ref{eqn:corrected_states}). With the ansatz
\begin{eqnarray}
	\psi_{\text{L}} & = & \cos\theta \, \tilde\psi_{+} + \sin\theta \, \tilde\psi_{-} \\
	\psi_{\text{R}} & = & \sin\theta \, \tilde\psi_{+} - \cos\theta \, \tilde\psi_{-} 
\end{eqnarray}
we minimize $\int_{-b}^0 |\psi_{\text{R}}|^2 dx + \int_0^b |\psi_{\text{L}}|^2 dx$ what is the probability to find a particle described by $\psi_{\text{R}}$ in the left bin and vice versa. One finds, that the best fitting for left and right states are given by
\begin{eqnarray}
	\psi_{\text{L}} & = &\frac{1}{\sqrt{2}} \left(\psi_{+} + \psi_{-}\right) \label{eqn:Lstate} \\
	\psi_{\text{R}} & = &\frac{1}{\sqrt{2}} \left(\psi_{+} - \psi_{-}\right) \label{eqn:Rstate} \, .
\end{eqnarray}
These are the same states as the optimal left and right states for the symmetric DWP. However, in this case $\theta$ reads
	\begin{equation}
		\theta \approx -\frac{\pi}{4}+\frac{m g h_0}{2\Delta E_n}
	\end{equation}
	and the action of the Hamiltonian is given by
\begin{align}
	(H_0 + V) \psi_{\text{L}} & = E_{\text{L}} \psi_{\text{L}} + K \psi_{\text{R}} \\
	(H_0 + V) \psi_{\text{R}} & = E_{\text{R}} \psi_{\text{R}} + K \psi_{\text{L}}
\end{align}
with the energies and the coupling constants  
\begin{align}
	E_{i,n,\text{R}/\text{L}} & = E_{i,n} \pm \frac12 \Delta E_n \cos(2\theta) \approx E_{i,n} \pm \frac12 m g h_0 \\
	K_n & = - \frac12 \Delta E_n \sin(2\theta) \approx \frac12 \Delta E_n \,.
\end{align}
We find in particular that the energy difference $E_{i,n,\text{R}} - E_{i,n,\text{L}} = m g h_0$ is related to the potential step \eqref{PotStep} and does not depend on $i$ or $n$.

The wave functions $\psi_{\text{L/R}}(x)$ are no eigenstates of the Hamilton operator. In order to check how the actual wave function decomposes into the left and right wave functions we make the ansatz
\begin{equation}
	\Psi(t,x,z) = \frac{1}{\sqrt{2}} \left[w_{\text{L}}(t) \psi_{\text{L}}(x,z) + w_{\text{R}}(t) \psi_{\text{R}}(x,z) \right]
\end{equation}
and determine the equations of motion for the pobabilities $w_{\text{L/R}}(t)$ from $i \hbar \dot{\Psi}(x,t) = [H_0 + V(x)] \Psi(x,t)$. 

\end{document}